\documentclass[10pt,conference]{IEEEtran}
\IEEEoverridecommandlockouts
\usepackage{cite}
\usepackage{amsmath,amssymb,amsfonts}
\usepackage{algorithmic}
\usepackage{graphicx}
\usepackage{textcomp}
\usepackage{xcolor}
\def\BibTeX{{\rm B\kern-.05em{\sc i\kern-.025em b}\kern-.08em
    T\kern-.1667em\lower.7ex\hbox{E}\kern-.125emX}}

\usepackage{xspace}
\usepackage{booktabs}
\usepackage{listings}
\usepackage[table]{xcolor}
\usepackage{threeparttable}
\usepackage{booktabs}

\usepackage{comment}

\newcommand{\plugin}{\textit{SAFE}\xspace} 

\usepackage[framemethod=TikZ]{mdframed}

\newcommand{\researchq}[1]{\textbf{RQ#1}\xspace} 

\newcommand\myshade{85}
\colorlet{mylinkcolor}{violet}
\colorlet{mycitecolor}{orange}
\colorlet{myurlcolor}{gray}

\usepackage{url}%
\usepackage{hyperref}
\usepackage{cleveref}

\hypersetup{
	linkcolor  = mylinkcolor!\myshade!black,
	citecolor  = mycitecolor!\myshade!black,
	urlcolor   = myurlcolor!\myshade!black,
	colorlinks = true,
}

\begin{document}

\title{Explaining Software Vulnerabilities with Large Language Models\\

}

\author{\IEEEauthorblockN{Oshando Johnson}
\IEEEauthorblockA{
\textit{Fraunhofer IEM}\\
Paderborn, Germany \\
oshando.johnson@iem.fraunhofer.de}
\and
\IEEEauthorblockN{Alexandra Fomina}
\IEEEauthorblockA{
\textit{Chapman University}\\
California, United States \\
fomina@chapman.edu}
\and
\IEEEauthorblockN{Ranjith Krishnamurthy}
\IEEEauthorblockA{
\textit{Fraunhofer IEM}\\
Paderborn, Germany \\
ranjith.krishnamurthy@iem.fraunhofer.de}
\and[\hfill\mbox{}\par\mbox{}\hfill]
\IEEEauthorblockN{Vaibhav Chaudhari}
\IEEEauthorblockA{
\textit{Paderborn University}\\
Paderborn, Germany \\
vaibhav.chaudhari@uni-paderborn.de}
\and
\IEEEauthorblockN{Rohith Kumar Shanmuganathan}
\IEEEauthorblockA{
\textit{University of Oldenburg}\\
Oldenburg, Germany \\
rohith.shanmuganathan@uol.de}
\and
\IEEEauthorblockN{Eric Bodden}
\IEEEauthorblockA{
\textit{Paderborn University and Fraunhofer IEM}\\
Paderborn, Germany \\
eric.bodden@uni-paderborn.de}
}

\maketitle

\begin{abstract}

The prevalence of security vulnerabilities has prompted companies to adopt static application security testing (SAST) tools for vulnerability detection.
Nevertheless, these tools frequently exhibit usability limitations, as their generic warning messages do not sufficiently communicate important information to developers, resulting in misunderstandings or oversight of critical findings.
In light of recent developments in Large Language Models (LLMs) and their text generation capabilities, our work investigates a hybrid approach that uses LLMs to tackle the SAST explainability challenges. In this paper, we present \plugin, an Integrated Development Environment (IDE) plugin that leverages GPT-4o to explain the causes, impacts, and mitigation strategies of vulnerabilities detected by SAST tools. Our expert user study findings indicate that the explanations generated by \plugin can significantly assist beginner to intermediate developers in understanding and addressing security vulnerabilities, thereby improving the overall usability of SAST tools. 

\end{abstract}

\begin{IEEEkeywords}
vulnerability explanation, static analysis, large language models, explainability, vulnerability detection
\end{IEEEkeywords}

\section{Introduction}
\label{sec:intro}

With the rise in software security vulnerabilities such as those in the Common Weakness Enumeration (CWE) Top 25 Most Dangerous Software Weaknesses list \cite{cwetop25}, many companies resort to static application security testing (SAST) tools for the detection of software vulnerabilities. 
Given that many SAST tools provide generic warning messages and also lack sufficient information about the detected vulnerabilities, developers often misunderstand or ignore the tool findings \cite{study_usability_sast}. As an alternative, developers have expressed the need for improved warning messages similar to the explanations found on blogs and online forums \cite{study_usability_sast}.

To address the explainability challenges of SAST tools, recent developments in Large Language Models (LLMs) have provided new possibilities for security-oriented tasks such as vulnerability detection, given the capability of LLMs to understand code and convey information in natural language \cite{llmvulexp}. However, these approaches often present LLMs as possible alternatives for established static application security testing approaches. 
Despite advancements in vulnerability detection using LLMs, previous studies have unequivocally demonstrated that for critical systems, SAST is advisable due to its ability to deliver the requisite reliability and precision \cite{gnieciak2025large}. This underscores the necessity of investigating complementary roles of LLMs in vulnerability detection rather than viewing them only as substitutes. However, there is little research exploring hybrid approaches that integrate SAST tools with LLMs to leverage the strengths of both methodologies.
To address this research gap, we explore the effectiveness of using LLMs to explain the cause, impact, and mitigation for vulnerabilities detected by SAST tools \cite{study_usability_sast}. 

In this paper, we present \plugin (Static Analysis Findings Explainer), an IntelliJ IDEA plugin that leverages LLMs to explain vulnerabilities for developers with limited software security experience. To identify the most suitable LLM for \plugin, we benchmarked widely used open and closed-source models on vulnerability detection tasks using various prompt engineering techniques. Using one of the top performing model, GPT-4o, we evaluated \plugin in an expert study, which revealed that the explanations are helpful for developers with beginner to intermediate experience in software security.

We present \plugin in Section \ref{sec:approach}, evaluate it in Section \ref{sec:evaluation}, review related work in Section \ref{sec:related-work}, and conclude in Section \ref{sec:conclusion}. The plugin source code, demo and evaluation results are available online\cite{plugin}. 

 \begin{figure*}[ht]
\centering
\includegraphics[width=1\textwidth]{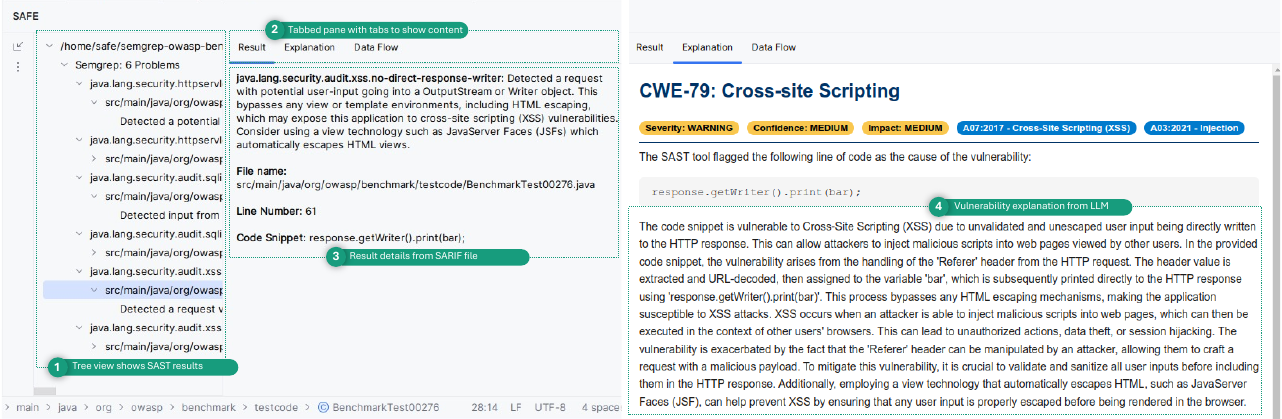}
\caption{\plugin's tool window screenshot showing the tree view (\includegraphics[width=0.014\textwidth]{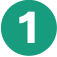}) and tabbed pane (\includegraphics[width=0.014\textwidth]{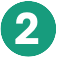}) containing
tabs for result details, explanations, and data-flow. The result details (\includegraphics[width=0.014\textwidth]{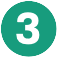}) and explanation (\includegraphics[width=0.014\textwidth]{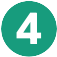}) for a sample cross-site scripting vulnerability are shown.}
\label{fig:user-interface}
\end{figure*} 

\section{Approach}
\label{sec:approach}

\begin{figure}[th!] 
\centering
\includegraphics[scale=0.90]{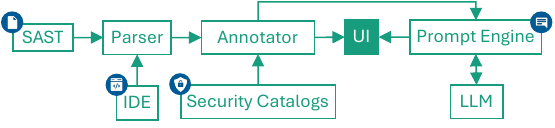}
\caption{Architecture of the \plugin Integrated Development Environment plugin for explaining static analysis tool results with large language models.}
\label{fig:architecture}
\end{figure}

In this section, we describe the \plugin plugin, which explains software vulnerabilities detected by SAST Tools. Figure \ref{fig:architecture} shows an overview of the plugin's architecture. 
The \textit{parser} processes SAST result files and extends the vulnerability findings with code snippets sourced from the \textit{Integrated Development Environment (IDE)}. Additionally, the results are enriched with supplementary information from \textit{security catalogs}~\cite{ide-workshop-paper} through the \textit{annotator}. The prompt engine utilizes the annotated results to construct a prompt, which is sent to the LLM to produce explanations. The plugin's user interface (UI) displays both the identified vulnerabilities and their corresponding explanations.
In the following sections, we describe the modules for parsing and annotating the SAST results, generating explanations, and the user interface.

\subsection{Parsing and Annotating SAST Vulnerability Results}
\label{subsec-result_parsing}
The parser processes the output of static analysis tools provided in the industry standard \emph{Static Analysis Results Interchange Format} (SARIF) \cite{sarif}. From the SARIF file, the parser extracts a property named \textit{results}, an array containing single results detected by the SAST tool \cite{sarif}. The property contains objects for the result's type, severity, message, and location. Using the \textit{location} property, the source code of the method containing the vulnerability is obtained using application programming interfaces (APIs) provided by the IntelliJ IDEA platform.
When available, the parser extracts the \textit{threadFlows} property, which is an array storing the code location for a specific single thread execution path through the program \cite{sarif}. From this property, the parser extracts data-flow information relevant for taint analysis, such as the source, intermediate elements, and sink for the vulnerability \cite{ide-workshop-paper}. The results are annotated with information from two security catalogs: the CWE list of software weaknesses \cite{cwetop25} and a list of critical methods \cite{ide-workshop-paper} in programs that influence security.

\subsection{Explaining SAST Results with Large Language Models}

To explain the results, SAFE employs a zero-shot prompting strategy \cite{kong2024better}, in which the explanation task is presented to the model without any prior examples. Using role-playing \cite{kong2024better}, we assign the role of a security expert to the LLM in the system prompt, while the user prompt outlines the specific explanation task. Using the data extracted in Sub-Section \ref{subsec-result_parsing} from the SARIF file, we populate the below zero-shot prompt template. 

\begin{mdframed}[linecolor=black,linewidth=1pt,
roundcorner=5pt,leftmargin=1, rightmargin=1 ,innertopmargin=7pt,innerbottommargin=7pt
]
{\fontsize{9pt}{11pt}\selectfont
\textit{\textbf{System}}: You are an assistant with expertise in explaining software security vulnerabilities in code snippets. You will be given a code snippet and the result from a static analysis security testing tool for the code snippet. Your task is to explain the static analysis result to a software developer based on their software security experience level. Provide information about the underlying cause, consequences, and mitigation strategies for the reported vulnerability.
\\When providing a response, follow these guidelines: \\\textit{\{Formatting and output Guidelines\}.}}
\end{mdframed}

\begin{mdframed}[linecolor=black,linewidth=1pt,
roundcorner=5pt,leftmargin=1, rightmargin=1 ,innertopmargin=7pt,innerbottommargin=7pt
]
{\fontsize{9pt}{11pt}\selectfont
\textit{\textbf{User}}: Explain the vulnerability detected in the code snippet to a developer who has \{\textit{level}\} experience in software security.
\\Detected Vulnerability: \{\textit{rule name}\}:\{\textit{rule message}\}
\\Code Snippet: \{\textit{location}\}
\\Line with vulnerability: \{\textit{location-line}\}
\\Data-flow: \{\textit{data-flow}\}
}
\end{mdframed}

The \textit{level} field is replaced with the user's experience level (beginner, intermediate, or advanced) with software security and impacts the level of detail of the explanations. The \textit{rule-name} and \textit{rule-message} fields are replaced with the vulnerability rule name and warning message reported by the SAST tool. The \textit{location} and \textit{location-line} placeholders are replaced with the source code of the vulnerable method and line, respectively. 
Finally, the \textit{data-flow} placeholder is replaced with the taint source, intermediate, and sink lines of codes.

\subsection{User Interface}

The plugin's user interface uses a tool window (child windows in IntelliJ Idea for displaying information) and contains a toolbar with tool window buttons. Figure \ref{fig:user-interface} displays a screenshot of the plugin's tool window with the result message and explanation for a cross-site scripting vulnerability detected by the static analysis tool Semgrep v1.119.0. A screencast of the plugin is available on GitHub \cite{plugin}.  
The tool window is divided into two parts: a custom tree view to hierarchically display the results from the SARIF file (\includegraphics[width=0.015\textwidth]{figures/1.png}) and a tabbed pane to show content (\includegraphics[width=0.015\textwidth]{figures/2.png}). The Results tab (\includegraphics[width=0.015\textwidth]{figures/3.png}) shows the result details (\textit{results} property), the Explanation tab shows the LLM explanation (\includegraphics[width=0.015\textwidth]{figures/4.png}), and the data-flow tab shows data-flow information from the \textit{threadflow} property. 
The explanation tab additionally contains the vulnerability type, severity, and impact, as well as the SAST tool's confidence, general mitigation strategies, and thumbs up/down feedback buttons.

\section{Evaluation}
\label{sec:evaluation}

In this section, we evaluate \plugin's main objective of explaining vulnerabilities detected by static analysis tools with LLMs using the following research questions:

\begin{itemize}
    \item \emph{\hyperref[rq:rq1]{\researchq{1}}}: Which large language models and prompt engineering techniques achieve the best performance in detecting security vulnerabilities?
    \item \emph{\hyperref[rq:rq2]{\researchq{2}}}: To what extent can large language models explain vulnerabilities detected by static analysis tools for users with different software security experience?
\end{itemize}

For the evaluation, we use the OWASP Benchmark \cite{owaspbench}, a comprehensive Java vulnerability test suite containing more than 2000 test cases across 11 vulnerability types. In the next subsections we describe the experiments and results.

\subsection{RQ1: Vulnerability Detection with Large Language Models}
\label{rq:rq1}

Before evaluating the quality of explanations provided by LLMs (addressed in \hyperref[rq:rq2]{RQ2}), it is essential to first determine which model and prompting strategy are most suitable for the task of vulnerability detection. To this end, we developed a benchmarking framework, VuLLMBench \cite{vullmbench}, that systematically compares multiple LLMs and different prompting techniques for vulnerability detection. The rationale is that if a model employing a specific prompting strategy excels in vulnerability detection, it is likely to exhibit internal reasoning capabilities, which may enable it to produce reliable explanations of the identified vulnerabilities.

We evaluated role-based prompts without examples (zero-shot), with code examples (few-shot), and also with a series of intermediate reasoning steps (chain-of-thought). Using these prompting techniques, we tested 22 open- and closed-source LLMs that have been recommended for vulnerability detection tasks. Our experiments confirmed previous research~\cite{chen2023unleashing}, which reported that zero-shot prompts are more effective for vulnerability detection tasks. 

To further evaluate the robustness, generalization, and real-world applicability of the models, we experimented with basic name-based obfuscation techniques in which variables, methods, and classes were renamed. Experimenting with obfuscated code provides insights into how well models can identify vulnerabilities when the code is altered, thereby testing their ability to discern underlying logic and security flaws. Table~\ref{table:llms_performance_owasp} reports the performance of the top 10 LLMs using a zero-shot prompt on the OWASP Benchmark for the original and obfuscated test cases. The complete results are available in the project repository \cite{vullmbench}.

\begin{table}[t]
    \centering
    \caption{Precision (P), recall (R) and F1-Score (F1) for vulnerability detection with LLMs on the OWASP Benchmark.}
    \label{table:llms_performance_owasp}
    \vspace{0.1in}
    \footnotesize
    \begin{tabular}{lcccccc}
        \toprule
     & \multicolumn{3}{c}{\textbf{\textit{Original}}}                                & \multicolumn{3}{c}{\textbf{\textit{Obfuscated}}}          \\ \cmidrule(lr){2-4} \cmidrule(lr){5-7}
    \textbf{Model } & \textbf{P} & \textbf{R} & \textbf{F1} & \textbf{P} & \textbf{R} & \textbf{F1} \\ 
    \midrule

GPT-5              & 0.78 & 0.98 & 0.87 & 0.75 & 0.96 & 0.84 \\
o3-mini            & 0.82 & 0.90 & 0.86 & 0.84 & 0.88 & 0.86 \\
GPT-5-mini         & 0.72 & 0.96 & 0.83 & 0.75 & 0.95 & 0.84 \\ 
GPT-4o             & 0.59 & 1.00 & 0.74 & 0.53 & 0.99 & 0.69 \\ 
Deepcoder (14B)    & 0.70 & 0.75 & 0.72 & 0.65 & 0.68 & 0.66 \\
Llama3.1 (70B)     & 0.57 & 0.96 & 0.72 & 0.56 & 0.96 & 0.70 \\
GPT-4o-mini        & 0.54 & 0.98 & 0.70 & 0.55 & 0.97 & 0.70 \\
Gemma2 (9B)        & 0.53 & 0.99 & 0.69 & 0.53 & 0.88 & 0.66 \\
CodeGemma (7B)     & 0.57 & 0.86 & 0.69 & 0.52 & 0.88 & 0.65 \\
CodeLlama (70B)    & 0.53 & 0.96 & 0.68 & 0.52 & 0.94 & 0.67 \\
    \bottomrule
    \end{tabular}
    \vspace{-12pt}
\end{table}

The reasoning models (GPT-5 and o3) outperformed the other models on the OWASP Benchmark with F1-Scores above 0.83, given their increased ability to make more reliable and accurate decisions, work through ambiguity that may exist in the code snippets, and solve problems \cite{reasoning}. The performance of these models could be possibly improved by designing prompts that following the guidelines for reasoning models such as using developer instead of system  messages \cite{reasoning}.
The general models (GPT-4o, Llama3.1, and GPT-4o-mini) as well as the code models (Deepcoder, CodeGemma and CodeLlama) also achieved relatively good F1-Scores due to high recall, however, the models have relatively low precision. 
These findings suggest that the models produce a high rate of false positives and are unable to effectively filter noise—a known challenge for developers using static analysis tools \cite{study_usability_sast}.

The performance of most of the models was negatively impacted due to the minor lexical changes from the name-based obfuscation. Although simply renaming identifiers does not impact the presence of the vulnerabilities and most SAST tools would be robust against such changes, some of the models seem to misled by such alterations in the code.

Although GPT-5 achieved the strongest overall performance in our benchmarks, its August 2025 release postdated our user study (see \hyperref[rq:rq2]{RQ2}), for which GPT-4o had already been selected; consequently, GPT-5 was not included. 
We added the reasoning model o3-mini to the benchmark alongside the GPT-5 reasoning models. 
Although o3-mini outperformed GPT-4o on the metrics in Table \ref{table:llms_performance_owasp}, it required approximately twice the runtime in our benchmarks and was more expensive. Accordingly, we selected GPT-4o, consistent with OpenAI’s guidance for coding and agentic tasks; moreover, GPT models remain well suited for well-defined tasks, with lower latency and cost \cite{reasoning}.

\subsection{RQ2: LLM-generated Explanations Evaluation}
\label{rq:rq2}

To evaluate the explanations, we analyzed the OWASP Benchmark with the open-source static analysis tool Semgrep (version 1.119.0). From the benchmark results, we selected two random samples of the top 3 most dangerous vulnerabilities \cite{cwetop25}, namely \textit{CWE22 Path Traversal}, \textit{CWE89 SQL Injection}, and \textit{CWE79 Cross-site Scripting}. Figure \ref{fig:user-interface} shows the message reported by Semgrep as well as SAFE's generated explanation. Semgrep’s message for the ``no-direct-response-writer" rule is intentionally generic, as it is reused across all detected instances. 
In contrast, \plugin's explanation provides a step-by-step walkthrough tailored to the specific finding, referencing the variables and lines of code that may contribute to the vulnerability. \plugin's  explanations align with research-recommended warning practices: they are detailed and descriptive, outline the analysis steps, and adapt to the coding context in which the vulnerability occurs \cite{8967437}.

To further assess the quality of \plugin's explanations, we conducted an expert study with four experienced software security trainers to evaluate the accuracy, correctness, readability, and helpfulness of the generated messages. Three of the trainers were certified scientific trainers. On average, they had four years of training experience and had delivered eight training sessions, primarily to participants with beginner to intermediate software security backgrounds. The trainings focused on secure software engineering and information technology security. We opted for a human evaluation study \cite{chiang-lee-2023-large} with experts to verify the explanations against the source code and the SAST tool’s results.

The one-hour study comprised a survey to capture participants’ training experience, a plugin demonstration, an expert usability test, and an interview. Participants evaluated the LLM-generated explanations using five criteria or attributes: 

\begin{enumerate}
    \item \textit{Relevant}: relates to vulnerability, appropriate for the user’s experience, fitting vocabulary, and essential details
    \item \textit{Faithful}: free from hallucination (information that is not supported by the source text and vulnerability)
    \item \textit{Concise}: information-dense, does not repeat the same point multiple times, and is not unnecessarily verbose 
    \item \textit{Coherent}: well-structured, easy to follow, not just a jumble of facts, grammatically and syntactically sound
    \item \textit{Accuracy}: captures the original warning message and code meaning
\end{enumerate}

Relevance and cohesiveness (coherency) are commonly used in human evaluations of generated text \cite{chiang-lee-2023-large}. Because effective warning messages should clearly identify the detected issue, explain why it matters, and describe how to fix it \cite{study_usability_sast}, we emphasized conciseness. For vulnerability detection tasks, it is critical to convey accurate information; accordingly, we assessed faithfulness and accuracy.
These criteria are evaluated using a 5-point Likert scale with options ranging from \textit{very poor} to \textit{very good}.

\begin{figure}[ht] 
    \vspace{-10pt}
\centering
\includegraphics[scale=0.59]{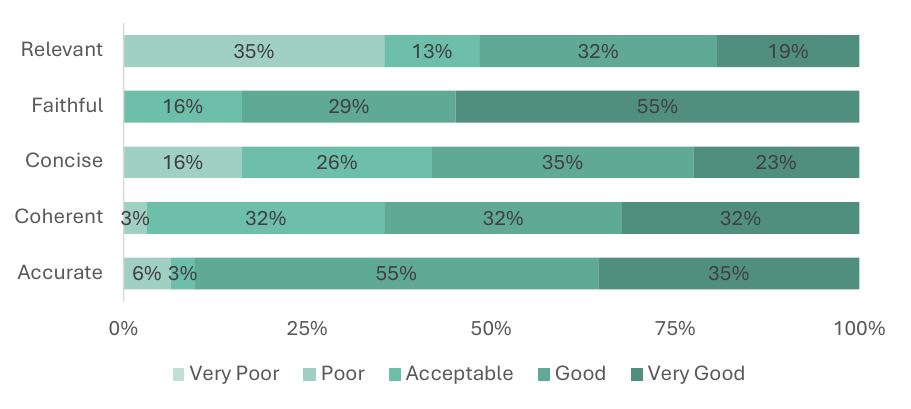}
\caption{Stacked bar chart showing the evaluation of the vulnerability explanations using a Likert scale with options from very poor to very good.}
\label{fig:expert-study}
\end{figure}

Figure \ref{fig:expert-study} shows the expert evaluations of the explanations for the vulnerabilities detected in the OWASP Benchmark with Semgrep. For all of the criteria, the trainers considered 64\% of the explanations to be at least \textit{acceptable} and none of them were considered to be \textit{very poor}. The \textit{relevant} criteria had the highest distribution (35\%) of \textit{poor} evaluations, arising from an overlap in the content of explanations for the different levels and explanations being more relevant for another level. Although the trainers mentioned that beginner and intermediate explanations were usually fitting, more general explanations would be helpful for beginners, while intermediate explanations could explain the detected vulnerability.

For the \textit{faithful} category, all of the explanations evaluated were considered to be at least good and therefore free from hallucinations. This was also confirmed in an initial assessment by four of the authors: the explanations contained no hallucinations and correctly referenced the given the prompt-provided context.
The distribution among the responses for the \textit{concise} criteria arises because the explanations contained information that was not fitting, repeated or unnecessary, too wordy, too technical, and often a mere rephrasing of the SAST tool output. A similar evaluation of the \textit{coherent} criteria is observed given the even split across \textit{acceptable, good and very good}. The trainers cited that the structure of the explanation was not always logical and missed important information such as mitigation strategies. 90\% of the explanations were considered to be \textit{accurate} (\textit{good and very good}) given that the code was often correctly explained by the LLM. However, in the initial assessment by the authors, it was observed that even for false positives, the explanations still implied a vulnerability in the source code, indicating that the language model tended to assume all findings were true positives. This issue can be mitigated by instructing the large language model to validate each reported vulnerability or by applying other false-positive detection methods before generating explanations.

Qualitatively, the participants found the generated explanations useful for novice and intermediate users and more effective than the SAST tool messages for explaining the vulnerability causes, impact, and mitigation strategies. They recommended improvements to explanation utility, automatic skill-level detection, and overall plugin usability improvements. Explanation quality (relevance, concision, coherence) could be enhanced by refining the prompt template with clearer instructions and sufficient task context. They also advocated for a clearer distinction between novice and intermediate levels and level-specific prompts, rather than a single prompt for all users.

\section{Related Work}
\label{sec:related-work}

Prior work on explainability in static analysis has introduced contextual information, visualizations, and standardized reporting formats. For example, rule graphs encode the rules and data flows underlying vulnerability detection, thereby visualizing the internal reasoning of analysis tools \cite{9106860}. While effective, this approach assumes domain knowledge of taint analysis to interpret the depicted flows.
Our approach addresses this barrier by using AI-generated natural-language explanations of data flows, making them accessible to developers with limited software security expertise.

In the domain of artificial intelligence, prior work has examined large language models for vulnerability detection and explanation, often positioning them as replacements for state-of-the-art AI techniques and static application security testing (SAST).
LLM-based approaches such as LLMVulExp \cite{llmvulexp} focus on using various prompt engineering and fine tuning techniques to enhance vulnerability detection, explanation and repair. LLM4SA \cite{auto_inspect_warnings_llm_how_far_are_we} builds on these approaches but also considers the static analysis warnings in order to filter out false positives.
Using LLMs to explain vulnerabilities detected by static analysis tools has not been well-researched and \plugin bridges this gap by providing explanations using the calling context in which the vulnerability was detected.

\section{Limitations and Threats to validity}

We acknowledge several threats to validity that may affect the interpretation and generalizability of our results, which we outline below.

\paragraph{External validity} Our evaluation relied on security experts rather than the target user population (beginner/intermediate developers), which may limit generalizability. We plan to refine the plugin based on expert feedback and conduct a large-scale user study with developers at these levels.

\paragraph{Construct/conclusion validity}: RQ1 underpins RQ2, but we did not analyze the relationship between vulnerability detection performance and explanation quality in LLMs. This warrants further investigation through experiments that compare multiple models, analyze their outputs to identify similarities and differences, and quantify the correlation between detection accuracy and explanatory quality. We plan to undertake a more comprehensive investigation of the relationship between LLMs’ vulnerability detection performance and explanatory capabilities.

\section{Conclusion}
\label{sec:conclusion}

In this paper, we address some of the usability/explainability limitations of static analysis tools by developing an approach \plugin that uses LLMs to explain security vulnerabilities. We bench-marked various models and prompt engineering techniques and evaluated the explanations generated with GPT-4o in a study with software security trainers. Our results showed that a hybrid approach combining LLMs with SAST can help to improve the limitations of both approaches for vulnerability detection tasks, especially for developers with beginner to intermediate software security experience.

\section*{Acknowledgment}

We acknowledge support by the German state of North Rhine-Westphalia for the CyberResilience.nrw Project funded through the NEXT.IN.NRW innovation competition.

\bibliographystyle{IEEEtran}
\bibliography{references}

\end{document}